\documentclass[twocolumn,showpacs,aps,showkeys,superscriptaddress,amsmath,amssymb,nofootinbib,floatfix,prc,PREPRINT]{revtex4-1}

\usepackage{CJK}
\usepackage{multirow}
\usepackage{graphicx}
\usepackage{color}
\usepackage{amsmath}
\usepackage{amssymb}
\usepackage[colorlinks=true,pdftex,bookmarks]{hyperref}
\usepackage[top=1in, bottom=1in, left=0.8in, right=0.8in]{geometry}
\usepackage{graphicx}

\newcommand{ \be }{\begin{eqnarray}}
\newcommand{ \ee }{\end{eqnarray}}

\usepackage{dcolumn} 
\usepackage{bm} 

\usepackage{graphicx}
\usepackage{color}
\definecolor{dgreen}{cmyk}{1.,0.,1.,0.4}        
\definecolor{orange}{cmyk}{0.,0.353,1.,0.}    

\usepackage[pdftex,bookmarks]{hyperref}

\usepackage{lineno}

\begin{document}


\begin{CJK*}{GB}{} 

\title{Hadronic ``flow" in p--Pb collisions at the Large Hadron Collider?}


\author{You Zhou}
\email{You Zhou: you.zhou@cern.ch}
\affiliation{Department of Physics and State Key Laboratory of Nuclear Physics and Technology, Peking University, Beijing 100871, China}
\affiliation{Niels Bohr Institute, University of Copenhagen, Blegdamsvej 17, 2100 Copenhagen, Denmark}
\author{Xiangrong Zhu}
\affiliation{Department of Physics and State Key Laboratory of Nuclear Physics and Technology, Peking University, Beijing 100871, China}
\author{Pengfei Li}
\affiliation{Department of Physics and State Key Laboratory of Nuclear Physics and Technology, Peking University, Beijing 100871, China}
\author{Huichao Song}
\email{Huichao Song: huichaosong@pku.edu.cn}
\affiliation{Department of Physics and State Key Laboratory of Nuclear Physics and Technology, Peking
University, Beijing 100871, China}
\affiliation{Collaborative Innovation Center of Quantum Matter, Beijing 100871, China}
\affiliation{Center for High Energy Physics, Peking University, Beijing
100871, China}

\date{\today}

\begin{abstract}
Using the Ultra-relativistic Quantum Molecular Dynamics ({\tt UrQMD}) model, we investigate azimuthal correlations in p--Pb collisions at $\sqrt{s_{_{\rm NN}}}=5.02$ TeV. Comparison with the experimental data shows that {\tt UrQMD} can not reproduce the multiplicity dependence of 2- and 4-particle cumulants, especially the transition from positive to negative values of $c_{2}\{4\}$ in high multiplicity events, which has been taken as experimental evidence of collectivity in p--Pb collisions.
Meanwhile, {\tt UrQMD} can not qualitatively describe the differential elliptic flow, $v_{2}(p_{\rm T})$, of all charged hadrons at various multiplicity classes. These discrepancies show that the simulated hadronic p--Pb systems can not generate enough collective flow as observed in experiment, the associated hadron emissions are largely influenced by non-flow effects. However, the characteristic $v_{2}(p_{\rm T})$ mass-ordering of pions, kaons and protons is observed in {\tt UrQMD}, which is the consequence of hadronic interactions and not necessarily associated with strong fluid-like expansions.
\end{abstract}


\maketitle

\end{CJK*}

\section{Introduction}
\label{s:Introduction}

The relativistic heavy ion collisions at Relativistic Heavy Ion Collider (RHIC) and the Large Hadron Collider (LHC) have provided strong evidences for the creation of the Quark--Gluon Plasma (QGP)~\cite{Rev-Arsene:2004fa,Gyulassy:2004vg,Muller:2006ee,Muller:2012zq}. One of the crucial observables is the azimuthal anisotropy of the transverse momentum distribution for produced hadrons~\cite{Ollitrault:1992bk}. As a signature of the collective flow, it provides important information on the Equation of State (EoS) and the transport properties of the QGP~\cite{reviews,Teaney:2009qa,Romatschke:2009im,Gale:2013da,Heinz:2013th,Song:2012ua}. Usually, the anisotropy is characterised by the Fourier flow-coefficients~\cite{Voloshin:1994mz}:
\begin{equation}
v_{n}=\langle\cos[n(\varphi-\Psi_{n})\rangle,
\end{equation}
\label{eq1}
where $\varphi$ is the azimuthal angle of the emitted hadrons, $\Psi_{n}$ is the $n^{\rm{th}}$--order participant (symmetry) plane angle and $\langle \rangle$ denotes an average of all particles in all events.
The second Fourier flow-coefficient $v_{2}$ is called elliptic flow, which is associated with the initial elliptic overlap region of the two colliding nuclei. In the past decades, most attention had been paid to the elliptic flow $v_{2}$, which has been systematically measured and studied at the Super Proton Synchrotron (SPS)~\cite{Alt:2003ab}, RHIC~\cite{Ackermann:2000tr,Adams:2004bi,Adcox:2002ms,Adamczyk:2013gw}, and the LHC~\cite{Aamodt:2010pa,Abelev:2014pua,ATLAS:2011ah} (for a summary, please also refer to~\cite{Voloshin:2008dg, Snellings:2011sz}). More recently, it was realised that the higher order flow-coefficients are equally important, which provide information on the fluctuating initial profiles of the created QGP~\cite{Alver:2010gr,Bhalerao:2011yg,Luzum:2013yya,Petersen:2010cw,Holopainen:2010gz,Pang:2012he,Teaney:2012ke,Qiu:2012uy,Schenke:2012hg,Retinskaya:2013gca,ALICE:2011ab,ATLAS:2012at,Chatrchyan:2012wg,Adamczyk:2013waa,Adare:2011tg}. 

The measurements of azimuthal correlations in $\sqrt{s_{_{\rm NN}}} =$ 5.02 TeV p--Pb collisions at the LHC were originally aimed to provide reference data for the high energy Pb--Pb collisions, especially on the cold nuclear matter effects. However, a large amount of unexpected collective behaviors have been discovered by the ALICE, ATLAS and CMS Collaborations.
For instance, a symmetric double ridge structure on both near- and away-side has been observed in high multiplicity p--Pb collisions by the ALICE Collaboration~\cite{Abelev:2012ola}. In addition, the CMS Collaboration has showed compatible results between multi-particle (including 4-, 6- and 8-particles) and all-particle correlations with Lee-Yang Zero's (LYZ)~\cite{Bhalerao:2003xf}, which corresponds to $v_{2}\{4\} \approx v_{2}\{6\} \approx v_{2}\{8\} \approx v_{2}\{{\rm LYZ}\}$~\cite{Wang:2014rja} (These results have also been confirmed by the ATLAS~\cite{Aad:2013fja} and ALICE Collaborations~\cite{Abelev:2014mda}). Recently, the measurements of azimuthal correlations have been extended to identified hadrons~\cite{ABELEV:2013wsa,Khachatryan:2014jra}. A $v_2$ mass-ordering feature, which says that the differential elliptic flow at low transverse momentum region monotonically increases with the decrease of hadron mass, has been observed among pions, kaons and protons in high multiplicity events~\cite{ABELEV:2013wsa}. Similarly, the CMS Collaboration found the mass-ordering between ${\rm K_{S}^{0}}$  and $\Lambda(\overline{\Lambda})$, which showed the $v_{2}$ of ${\rm K_{S}^{0}}$ is larger than the one of $\Lambda(\overline{\Lambda})$ at lower $p_T$, followed by a crossing at $p_{\rm T}\sim 2 \ \mathrm{GeV}$~\cite{Khachatryan:2014jra}.
Many of these experimental measurements have been semi-quantitatively described by (3+1)-d hydrodynamic simulations from several groups~\cite{Bozek:2011if,Bozek:2013ska,Bzdak:2013zma,Qin:2013bha,Werner:2013ipa}, which supports the experimental claim that large collective flow has been developed in small p--Pb systems.

In Au--Au or Pb--Pb collisions at RHIC and the LHC, the collective flow mainly develops in the QGP phase since the QGP fireball has long enough lifetime to develop the momentum anisotropy until the saturation is almost reached~\cite{reviews,Kolb:2000sd,Song:2007ux}. Meanwhile, a certain amount of collective flow is further accumulated in the hadronic stage through the microscopic rescatterings, which leads to the $v_2$ mass-ordering among various hadron species~\cite{Nonaka:2006yn,Hirano:2007ei,Song:2010aq,Song:2013qma}. Compared with Au--Au or Pb--Pb collisions, the smaller systems created in p--Pb collisions have much shorter lifetime. As a result, the momentum anisotropy is not likely to reach saturation even if the QGP has been created. The measured azimuthal correlations in p--Pb collisions might be largely influenced by the hadronic evolution. On the other hand, non-flow effects (e.g. from hadron resonance decays) are significantly enhanced for a smaller system with much lower particle yields, which also contribute to 2-particle correlations~\cite{Voloshin:2008dg}.

With an assumption of early thermalization for the created p--Pb systems, hydrodynamics simulates the evolution of both QGP and hadronic phases, and associate the azimuthal correlations of all charge and identified hadrons with the collective expansion of the systems~\cite{Bozek:2011if,Bozek:2013ska,Bzdak:2013zma,Qin:2013bha,Werner:2013ipa}. In this paper, we assume that the high energy p--Pb collisions do not reach the threshhold of the QGP formation, only pure hadronic systems are produced. We utilize a hadron cascade model Ultra-relativistic Quantum Molecular Dynamics ({\tt UrQMD})~\cite{Bass:1998ca,Bleicher:1999xi,Petersen:2008kb} to simulate the evolution of the hadronic matter and then study the azimuthal correlations of final produced hadrons. Our research focuses on two aspects: (1) investigating whether pure hadronic interactions could generate the observed flow signatures in high multiplicity events; (2) studying the mass-ordering of 2-particle correlations in pure hadronic p--Pb systems.

The paper is organized as follows. Section~\ref{s:UrQMD} briefly introduces the {\tt UrQMD} model. Section~\ref{s:Analysis_Method} outlines the 2- and 4-particle Q-Cumulant method.  Section~\ref{s:RandD} compares experimental measurements with the {\tt UrQMD} calculations on 2- and 4- particle azimuthal correlations, including centrality dependence and transverse momentum dependence. Section~\ref{s:summary} summarizes and concludes this work.

\section{$\mathrm{\mathbf{UrQMD}}$ hadron cascade model}
\label{s:UrQMD}

{\tt UrQMD} is a microscopic transport model to describe hadron--hadron, hadron--nucleus and nucleus--nucleus collisions at relativistic energies, based on the Boltzmann equations for various hadron species~\cite{Bass:1998ca,Bleicher:1999xi,Petersen:2008kb}. It has successfully described the soft physics at the AGS and SPS energies, where the created systems are dominated by strongly interacting hadrons.

In {\tt UrQMD}, the initial hadron productions are modeled via the excitation and fragmentation of strings. For higher collision energies above $\sqrt{s_{_{\rm NN}}}=$ 10 GeV, the PYTHIA mode~\cite{Sjostrand:2003wg} is implemented to describe the hard processes and the related hadron productions. The classical trajectories of the produced hadrons are then simulated through solving a large set of Boltzmann equations with flavour dependent cross sections. In the later version, {\tt UrQMD} contains 55 baryon and 32 meson species with masses up to 2.25 GeV, supplemented by the corresponding antiparticles and isospin-projected states~\cite{Petersen:2008kb}.  The elementary cross sections in the collision terms are either fitted from the experimental data or calculated via models e.g. a modified additive quark model (AQM). For two closely propagating hadrons, whether or not a collision happens is determined by a critical distance associated with the related cross section. When all elastic and inelastic collisions cease and all unstable hadrons have decayed into stable hadrons, the system is considered to reach kinetic freeze-out. {\tt UrQMD} then outputs the momentum and position information of the final produced hadrons.

\begin{table}
  \centering
  \begin{tabular}[t]{c|c}\hline
Event class & $N_{ch}$ ($|\eta|<$ 1.0) \\
\hline
0-5\% &  $>$72 \\
5-10\% & 60-72 \\
10-20\% & 47-60 \\
20-40\% &  38-47\\
40-60\% &  17-23\\
60-100\% & $<$17\\
\hline
\end{tabular}
  \caption{Event class determination in {\tt UrQMD} according to the number of all charged hadrons within $|\eta|<$ 1.0.}\label{table1}
\end{table}

\begin{table}
  \centering
  \begin{tabular}[t]{c|c}\hline
Event class & $N_{ch}$ $(2.8<\eta<5.1)$\\
\hline
0-20\% & $>$88\\
20-40\% & 54-88\\
40-60\% & 30-54\\
60-80\% & 13-30\\
80-100\% & $\le$ 13\\
\hline
\end{tabular}
  \caption{Event class determination in {\tt UrQMD} according to the number of all charged hadrons within 2.8 $<\eta<$ 5.1.}\label{table2}
\end{table}

\begin{figure}[tb]
\begin{center}
\includegraphics[width=0.47\textwidth]{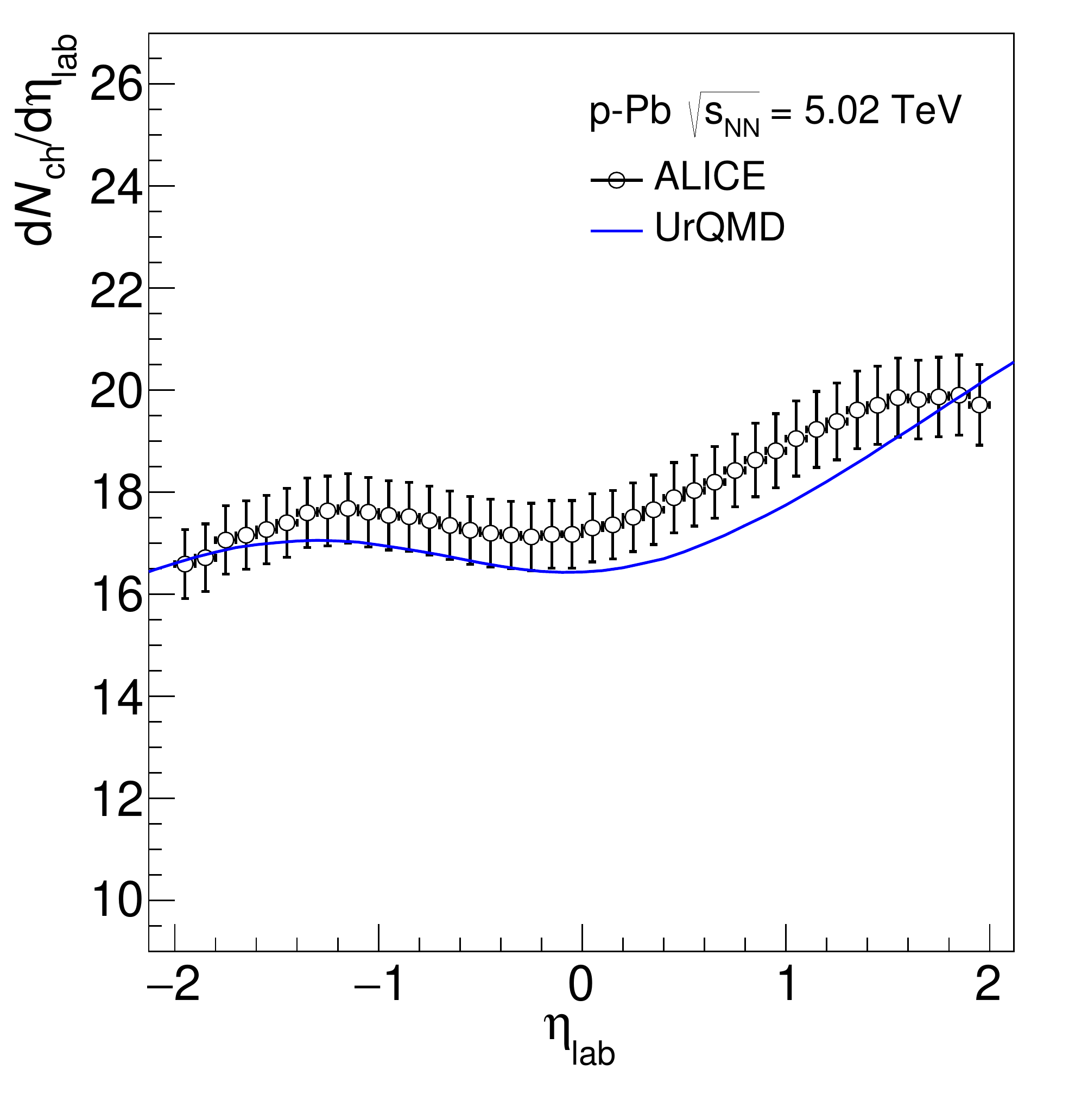}
\caption{Pseudorapidity density of all charged hadrons in minimum bias p--Pb collisions at $\sqrt{s_{_{\rm NN}}} =$ 5.02 TeV, measured by ALICE~\cite{ALICE:2012xs} and calculated from {\tt UrQMD}. }
\label{dndeta}
\end{center}
\end{figure}

In this paper, we implement {\tt UrQMD} version 3.4 to simulate the evolution of the assumed hadronic systems created in high energy p--Pb collisions. The simulations are executed in the equal speed system of two colliding nucleons with $\sqrt{s_{_{\rm NN}}}$ = 5.02 TeV. Correspondingly, the output information for final produced hadrons are defined in the centre-of-mass frame. In order to compare with the experimental data in the laboratory frame, we make a transformation between the centre-of-mass frame and the laboratory frame, which shifts the rapidity by 0.465. Following the related experimental papers~\cite{Abelev:2014mda,ABELEV:2013wsa,Abelev:2013haa}, the {\tt UrQMD} outputs are divided into several multiplicity classes, determined by the number of all charged hadrons $N_{ch}$ within a pseudorapidity range $|\eta| <$ 1 or 2.8 $< \eta <$ 5.1. The $N_{ch}$ values in these two centrality definitions are shown Table~\ref{table1} and~\ref{table2}.  The pseudorapidity density of all charged hadrons as a function of pseudorapidity in minimum bias p--Pb collisions is presented in Fig.~\ref{dndeta}.

\begin{figure*}[tb]
\includegraphics[width=0.95\textwidth]{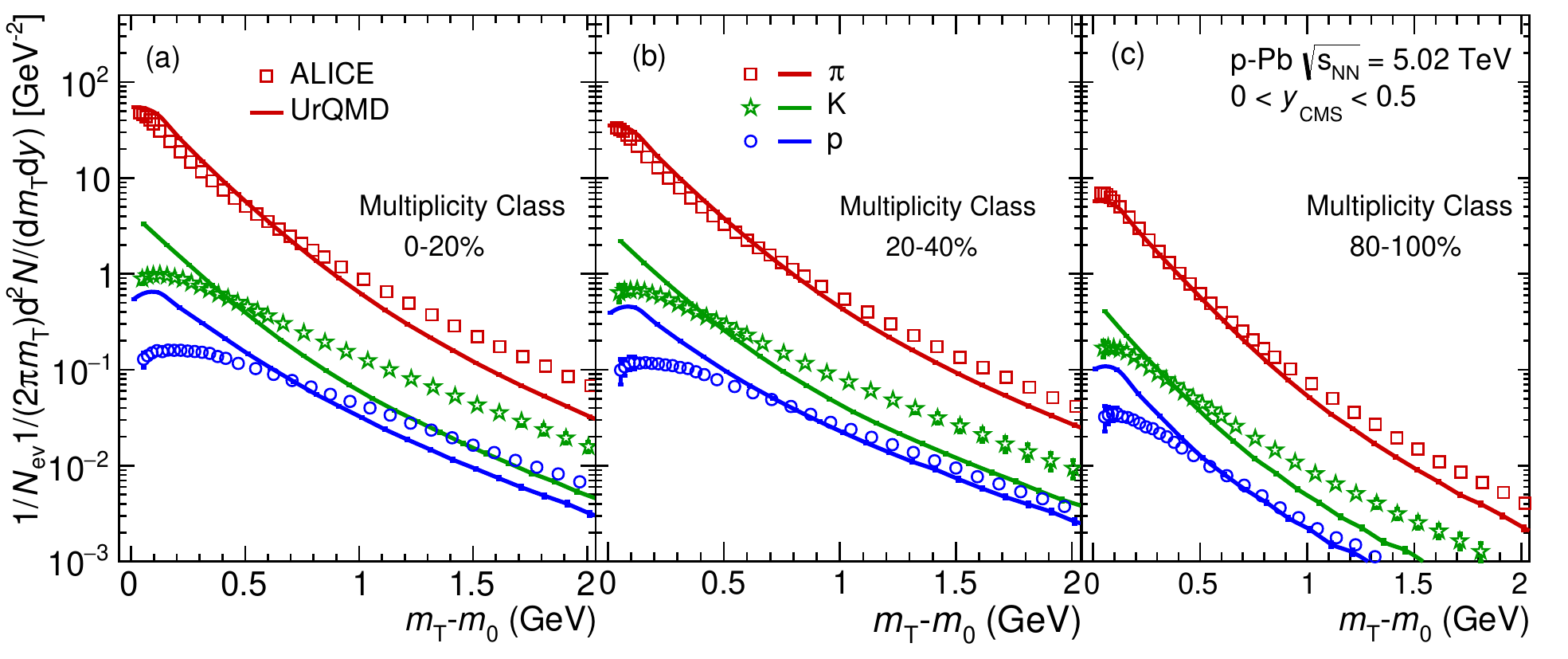}
\caption{$m_{\rm T}$ spectra of pions, kaons and protons in p--Pb collisions at $\sqrt{s_{_{\rm NN}}} =$ 5.02 TeV measured by ALICE~\cite{Abelev:2013haa} and calculated from {\tt UrQMD}. Here the multiplicity class determination in {\tt UrQMD} is based on Table~\ref{table2} .}
\label{ptSpectra}
\end{figure*}

\section{Analysis Method and Definitions}
\label{s:Analysis_Method}

In this paper, the azimuthal correlations are calculated using 2- and 4-particle Q-Cumulant method~\cite{Bilandzic:2010jr, Bilandzic:2013kga}, which were used in experiment at RHIC~\cite{Agakishiev:2011eq} and the LHC~\cite{Aamodt:2010pa, ALICE:2011ab, Wang:2014rja, Zhou:2014bba}.
In this method, both 2- and multi-particle azimuthal correlations are analytically expressed in terms of a Q-vector, which is defined as:
 \begin{equation}
Q_{n} = \sum_{i=1}^{M}e^{in\varphi_{i}},
\end{equation}
where $M$ is the multiplicity of the Reference Flow Particles (RFPs) and $\varphi$ is their azimuthal angle.
The single-event average 2- and 4-particle azimuthal correlations can be calculated via:
\begin{equation}
\begin{aligned}
\langle 2 \rangle & = \frac{|Q_{n}|^{2} - M} {M(M-1)}, \\
\langle 4 \rangle & = \frac{|Q_{n}|^{4} + |Q_{2n}|^{2} - 2 \cdot {\rm{Re}}[Q_{2n}Q_{n}^{*}Q_{n}^{*}]  } { M(M-1)(M-2)(M-3) }  \\
& ~ ~ - 2 \frac{ 2(M-2) \cdot |Q_{n}|^{2} - M(M-3) } { M(M-1)(M-2) },
\end{aligned}
\label{Eq:Mean24}
\end{equation}
here $\langle~\rangle$ stands for the average over all particles in a single event.

The 2- and 4-particle cumulants could be achieved as:
\begin{equation}
\begin{aligned}
c_{n}\{2\} & = \langle \langle 2 \rangle \rangle, \\
c_{n}\{4\} & = \langle \langle 4 \rangle \rangle - 2 \cdot \langle \langle 2 \rangle \rangle^{2},
\end{aligned}
\label{Eq:c24}
\end{equation}
here $\langle\langle~\rangle\rangle$ denotes the average over all particles over all events.

In order to proceed with the calculation of the differential flow of the Particles Of Interests (POIs), the $p_{n}$ and $q_{n}$ vectors for specific kinematic range and/or for specific hadron species are needed:
\begin{equation}
\begin{aligned}
p_{n}& = \sum_{i=1}^{m_{p}} e^{in\phi_{i}},\\
q_{n}& = \sum_{i=1}^{m_{q}} e^{in\phi_{i}},
\label{Eqpvector}
\end{aligned}
\end{equation}
where $m_{p}$ is the total number of particles labeled as POIs, $m_{q}$ is the total number of particles tagged both as RFP and POI.

The  single-event average differential 2- and 4-particle azimuthal cumulants are calculated as:
\begin{equation}
\begin{aligned}
\langle 2^{'} \rangle &= \frac{p_{n} Q_{n}^{*} - m_{q}} {m_{p}M - m_{q}},  \\
\langle 4^{'} \rangle  & = [p_{n}Q_{n}Q_{n}^{*}Q_{n}^{*}  - q_{2n}Q_{n}^{*}Q_{n}^{*} - p_{n}Q_{n}Q_{2n}^{*} \\
& - 2\cdot M p_{n} Q_{n}^{*} - 2 \cdot m_{q} |Q_{n}|^{2}  + 7 \cdot q_{n}Q_{n}^{*} - Q_{n}q_{n}^{*} \\
& + q_{2n}Q_{2n}^{*} + 2 \cdot p_{n} Q_{n}^{*} + 2 \cdot m_{q}M - 6 \cdot m_{q} ] .\\
&  / [ (m_{p}M - 3 m_{q})(M-1) (M-2) (M-3) ]
\end{aligned}
\label{Eq:Mean24p}
\end{equation}

For detectors with uniform azimuthal acceptance the differential 2- and 4-particle cumulants are given by:
\begin{equation}
\begin{aligned}
d_{n}\{2\} & = \langle \langle 2^{'} \rangle \rangle, \\
d_{n}\{4\} & = \langle \langle 4^{'} \rangle \rangle  - 2 \langle \langle 2^{'}\rangle \rangle \langle \langle 2\rangle \rangle.
\label{Eqdn24}
\end{aligned}
\end{equation}

Finally the estimated differential flow $v_{2}(p_{\rm T})$ from 2- and 4-particle correlations are given by:
\begin{equation}
\begin{aligned}
v_{n}\{2\}(p_{\rm T}) & = \frac{d_{n}\{2\}}{\sqrt{c_{n}\{2\}}},  \\
v_{n}\{4\}(p_{\rm T}) & = -\frac{d_{n}\{4\}}{( - c_{n}\{4\})^{3/4} }.
\end{aligned}
\end{equation}

Unfortunately, the $v_{n}$ obtained from the 2-particle Q-Cumulant contains contributions from so-called non-flow effects, which are additional azimuthal correlations between the particles due to e.g.
resonance decays, jet fragmentation, and Bose-Einstein correlations.
They can be suppressed by appropriate kinematic cuts. For instance, one can introduce a pseudorapidity gap between the particles in the 2-particle Q-Cumulant method~\cite{Zhou:2014bba}.
Accordingly, the whole event is divided into two sub-events, $A$ and $B$, which are separated by a $|\Delta\eta|$ gap.
This modifies Eq.~(\ref{Eq:Mean24}) to:
\begin{equation}
\langle 2 \rangle _{\Delta \eta} = \frac{Q_{n}^{A} \cdot Q_{n}^{B *}} {M_{A} \cdot M_{B}},
\label{Eq:Mean2Gap}
\end{equation}
where $Q_{n}^{A}$ and $Q_{n}^{B}$ are the flow vectors from sub-event $A$ and $B$,  $M_{A}$ and $M_{B}$ are
the corresponding multiplicities.

The 2-particle Q-Cumulant with a $|\Delta\eta|$ gap is given by:
\begin{equation}
c_{n}\{2, |\Delta\eta|\} = \langle \langle 2 \rangle \rangle _{\Delta\eta} .
\label{Eq:v22Gap10}
\end{equation}

For the calculations of differential flow with a pseudorapidity gap, there is no overlap of POIs and RPs if we select RPs from one subevent and POIs from the other.
This modifies Eqs.~(\ref{Eq:Mean24p}) to:
\begin{equation}
\langle 2^{'} \rangle_{\Delta \eta}  = \frac{p_{n,A} Q_{n,B}^{*} } {m_{p,A}M_{B}},
\label{Eqmean2pGap}
\end{equation}
and we get the differential 2-particle cumulant as:
\begin{equation}
d_{n}\{2, |\Delta\eta|\}  = \langle \langle 2^{'} \rangle \rangle_{\Delta\eta}.
\label{dn2Gap}
\end{equation}

Finally, the differential flow from 2-particle cumulant can be obtained by inserting the 2-particle reference flow (with $\eta$ gap) to the differential 2-particle cumulant:
\begin{equation}
v_{n}\{2, |\Delta\eta| \}(p_{\rm T})  = \frac{d_{n}\{2, |\Delta\eta|\}}{\sqrt{c_{n}\{2,|\Delta\eta|\}}}.
\label{vn2EtaGap}
\end{equation}

In this paper, the second and third Fourier flow-coefficients are evaluated using above equations, by setting the $n=$ 2 and 3, respectively.

\section{Results and Discussions}
\label{s:RandD}

\begin{figure*}[tbh]
\begin{center}
\includegraphics[width=0.9\textwidth]{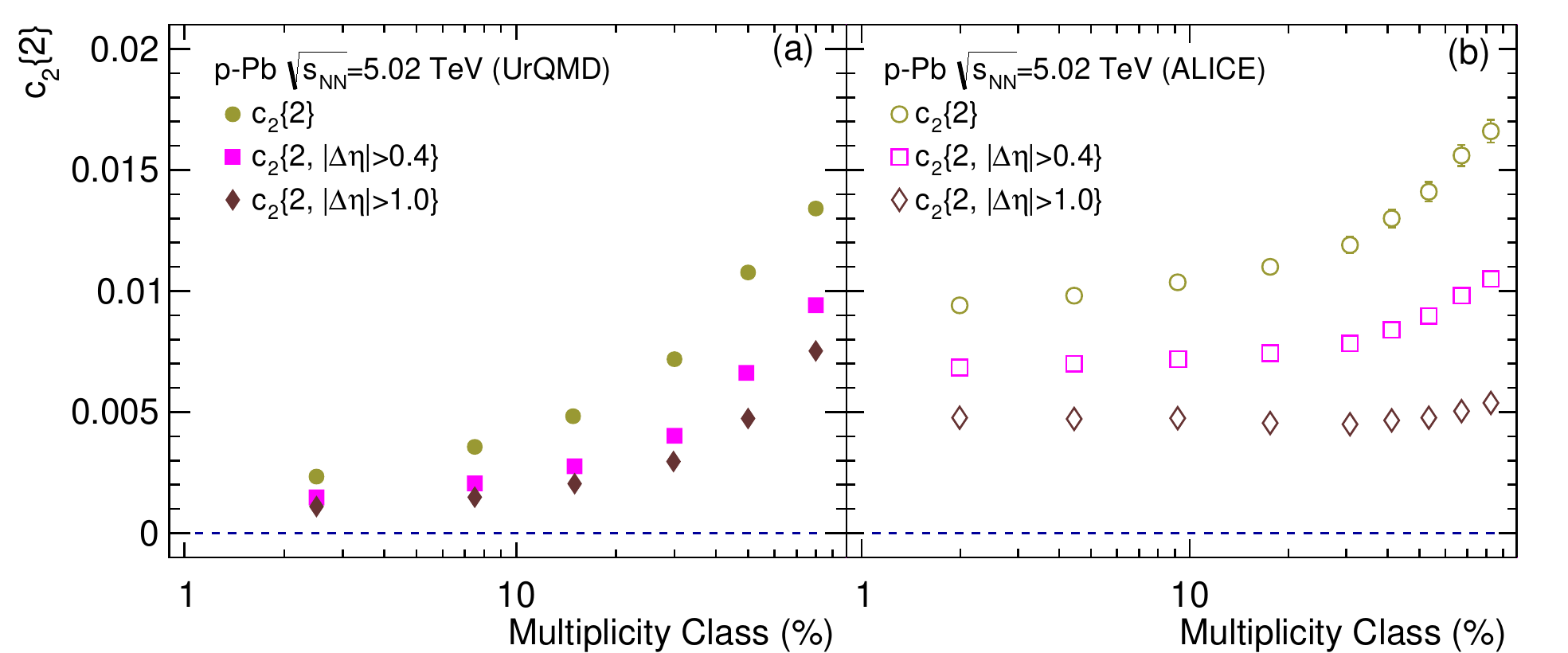}
\caption{(Color online) $c_{2}\{2\}$ of all charged hadrons in p--Pb collisions at $\sqrt{s_{_{\rm NN}}} =$ 5.02 TeV, calculated from {\tt UrQMD} (Left) and measured by ALICE (Right)~\cite{Abelev:2014mda}. The circle, square and diamond markers represent various pseudorapidity gap cuts without $\eta$ gap, with gap $|\Delta\eta|>$ 0.4 and $|\Delta\eta|>$ 1.0, respectively. Here the multiplicity class determination in {\tt UrQMD} is based on Table~\ref{table1}.}
\label{figure:c22}
 \end{center}
\end{figure*}

\begin{figure}
\centering
\includegraphics[width=0.48\textwidth,height=6.2cm]{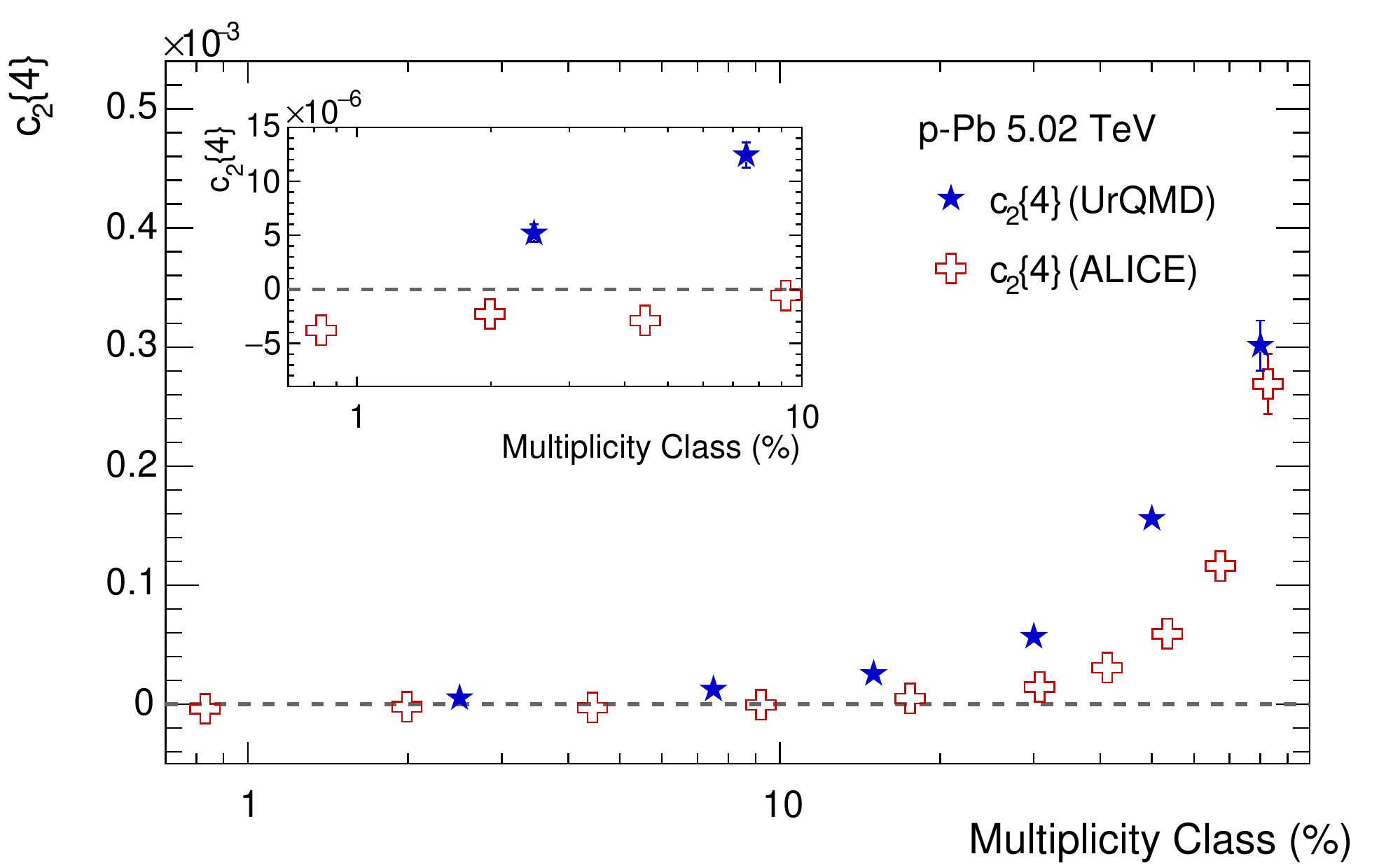}
\caption{(Color online) $c_{2}\{4\}$ of charged particles in p--Pb collisions at $\sqrt{s_{_{\rm NN}}} =$ 5.02 TeV, calculated from {\tt UrQMD} and measured by ALICE~\cite{Abelev:2014mda}. Here the multiplicity class determination in {\tt UrQMD} is based on Table~\ref{table1}.}
\label{figure:c24}
\end{figure}

\begin{figure*}[tbh]
\begin{center}
\includegraphics[width=0.9\textwidth]{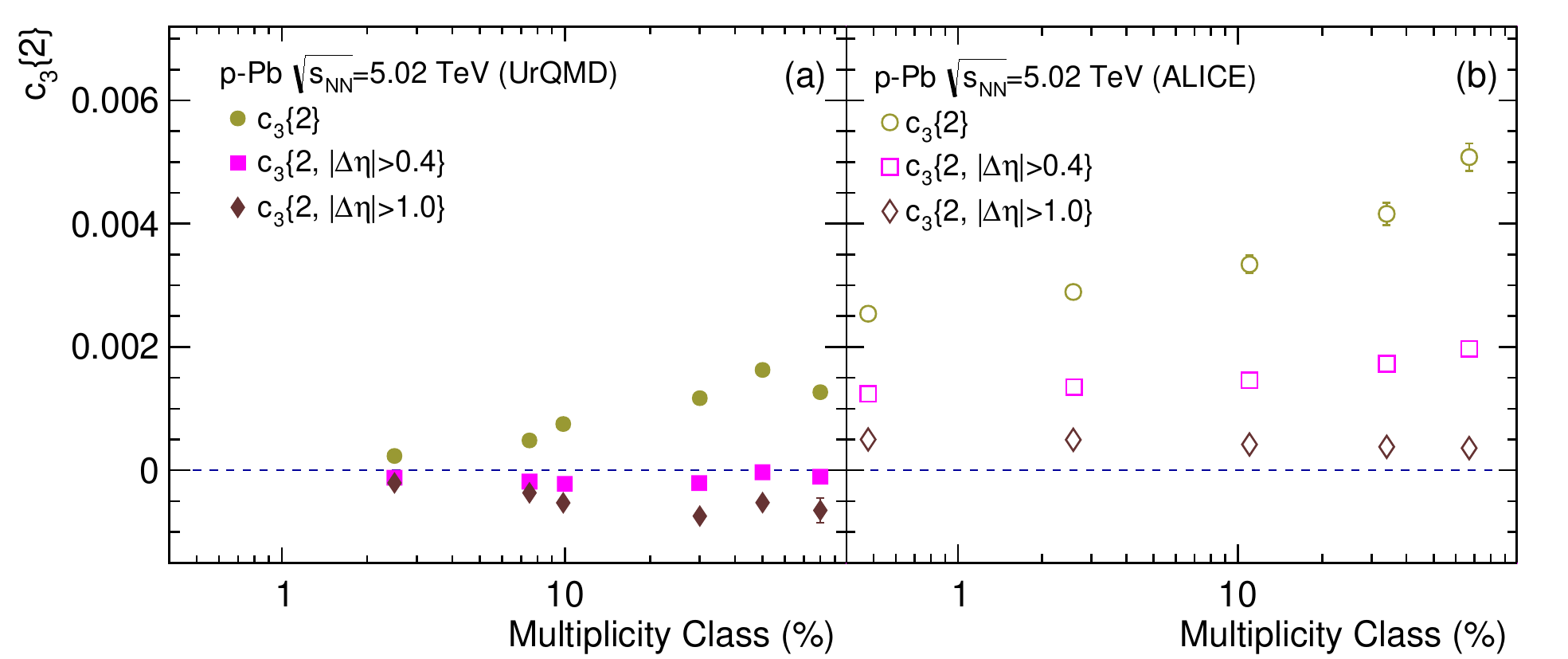}
\caption{(Color online) $c_{3}\{2\}$ of all charged hadrons in p--Pb collisions at $\sqrt{s_{_{\rm NN}}} =$ 5.02 TeV, calculated from {\tt UrQMD}(Left) and measured by ALICE (Right)~\cite{Abelev:2014mda}. Here the multiplicity class determination in {\tt UrQMD} is based on Table~\ref{table1}.}
\label{fig:c3234}
 \end{center}
\end{figure*}

This section mainly investigates the second and third Fourier flow-coefficients with cumulants in p--Pb collisions at $\sqrt{s_{_{\rm NN}}} =$ 5.02 TeV. Before studying the 2-particle and 4-particle correlations, it is important to check the single hadron information.
Fig.~\ref{dndeta} plots the pseudorapidity density of all charged hadrons $dN_{ch}/d\eta$ in minimum bias p--Pb collisions. In general, {\tt UrQMD} roughly describes the forward-backward asymmetry of the $dN_{ch}/d\eta$ curve within $|\eta|<2$. At midrapidity, $dN_{ch}/d\eta$ from {\tt UrQMD} are close to the ALICE measurements, but about 5\% lower than the experimental values.

Figure~\ref{ptSpectra} plots the $m_{\rm T}$ spectra of pions, kaons and protons in high energy p--Pb collisions. It is generally believed that, in the absence of radial flow, $m_{\rm T}$ spectra as a function of $m_{\rm T} - m_0$ ($m_0$ stands for the mass of the hadron and $m_{\rm T}=\sqrt{p_T^2 + m_0^2}$) satisfies the $m_{\rm T}$ scaling , where the slope of the spectra is independent of hadron species~\cite{Heinz:2004qz}. Such $m_{\rm T}$ scaling has been observed in p-p collisions at $\sqrt{s_{_{\rm NN}}} =$ 200 GeV~\cite{Heinz:2004qz,Barannikova:2002qw}.  In heavy ion collisions at the SPS energies and above, the $m_{\rm T}$ scaling is broken, which provides evidence for the development of strong radial flow in the hot QCD systems~\cite{Heinz:2004qz,Barannikova:2002qw,Afanasiev:2002fk,Velkovska:2001xz,Adler:2001aq}.

In high energy p--Pb collisions at $\sqrt{s_{_{\rm NN}}} =$ 5.02 TeV, the ALICE measurements in Fig.~\ref{ptSpectra} show that $m_{\rm T}$ scaling is broken at 0-20\% and 20-40\% multiplicity classes where the measured protons spectra are flatter than kaons ones~\footnote{The pion spectra is largely influenced by resonance decays at lower $m_{\rm T} - m_0$, which break the pion's $m_{\rm T}$ scaling even for the case without radial flow}. This provides evidence for the development of radial flow in high multiplicity events. The {\tt UrQMD} calculations in Fig.~\ref{ptSpectra} also present a weak broken of the $m_{\rm T}$ scaling, but show steeper spectra for pions, kaons and protons when compared with the ALICE curves. This indicates that the assumed hadronic p-Pb systems could not accumulate sufficient radial flow as observed in experiment.

With brief investigations of the single hadron data, we now focus on studying azimuthal correlations in high energy p--Pb collisions. Figure~\ref{figure:c22}  presents the centrality dependence of the 2-particle cumulant of the second Fourier flow-coefficient $c_{2}\{2\}$, calculated from  {\tt UrQMD} hadron cascade model (left) and measured by the ALICE collaboration (right). For various pseudorapidity gaps, $c_{2}\{2\}$ from {\tt UrQMD} exhibit decreasing trend from peripheral (low multiplicity events) to central collisions (high multiplicity events),  which agrees with the expectation of the azimuthal correlations not associated with the symmetry plane, the so-called non-flow effects. As the pseudorapidity gap increases, the magnitudes of $c_{2}\{2\}$  become weaker for both ALICE and {\tt UrQMD},  which illustrates that non-flow effects, usually few-particle correlations from resonance decays and jets, are suppressed by a large pseudorapidity gap. When the pseudorapidity gap $|\Delta \eta|$ is larger than 1.0, $c_{2}\{2\}$ from ALICE show much weaker centrality dependence, which is suggested as one of the hints for collective expansion in the created p--Pb systems. However, $c_{2}\{2\}$ from {\tt UrQMD} still present strong centrality dependence for $|\Delta \eta| >$ 1.0, showing a typical non-flow behavior. Usually, the non-flow effects between 2-particle correlations, denoted as $\delta_{n}$, behave as $\delta_{n} \sim 1/M$ where M is the multiplicity. The decreasing trend of $c_{2}\{2\}$ with the increase of multiplicity indicates that {\tt UrQMD} hadronic expansion could not generate enough flow in a small p-Pb system, non-flow effects are still pretty large even for the case with a large pseudorapidity gap cut $|\Delta \eta| >$ 1.0.

To better understand the hadronic systems simulated by {\tt UrQMD}, we investigate the 4-particle cumulant of the second Fourier flow-coefficient $c_{2}\{4\}$, which is equal to $-v_{2}\{4\}^{4}$  and expected to be less sensitive to non-flow effects.
Figure~\ref{figure:c24} plots the centrality dependence of $c_{2}\{4\}$ of all charged hadrons in p--Pb collisions at $\sqrt{s_{_{\rm NN}}} =$ 5.02 TeV.  Both the  {\tt UrQMD} and ALICE results show that $c_{2}\{4\}$ increase with the decrease of multiplicity from semi-central to peripheral collisions. For the most central collisions ($<$10\%),  $c_{2}\{4\}$ from ALICE  exhibits a transition from positive to negative values, indicating the creation of flow-dominated systems in the high multiplicity events. However, $c_{2}\{4\}$ from {\tt UrQMD} keeps positive for all available multiplicity classes, including the most central collisions. As a result, real values of $v_{2}\{4\}$ can not be extracted in {\tt UrQMD} for all centrality bins. This comparison further illustrates the difference between the p--Pb systems created in experiment and simulated by {\tt UrQMD}. The hadron emissions from {\tt UrQMD} are largely influenced by non-flow effects. Without the contributions from the initial stage and/or the QGP phase, the measured flow-like 4-particle correlations in high multiplicity events can not be reproduced by a microscopic transport model with only hadronic scatterings and decays.

In Figure~\ref{fig:c3234}, we further study the 2-particle azimuthal correlations for the third Fourier flow-coefficient $c_{3}\{2\}$. The {\tt UrQMD} calculations and the ALICE measurements with various pseudorapidity gaps are respectively shown in the left and right panels of Fig.~5. Similar to $c_{2}\{2\}$ in Fig.~\ref{figure:c22}, $c_{3}\{2\}$ also decreases with the increase of $|\Delta\eta|$. For the ALICE measurement, $c_{3}\{2\}$ keeps positive for all pseudorapidity gaps, which leads to real values of triangular flow $v_{3}\{2\}$ ($v_{3}\{2\}=\sqrt{c_{3}\{2\}}$) as measured in~\cite{Abelev:2014mda}.  Considered that non-flow effects are largely suppressed by a large pseudorapidity gap, the measured $c_{3}\{2\}$ at $|\Delta\eta|>1.0$ (and the associated triangular flow $v_{3}\{2\}$) is possibly mainly caused by collective expansion and reflects initial state fluctuations of the p-Pb systems. In contrast, $c_{3}\{2\}$ from {\tt UrQMD} turns to negative for $|\Delta\eta|>0.4$ and $|\Delta\eta|>1.0$, which does not produce a real value of $v_{3}\{2\}$  \footnote{We also find that $c_{3}\{4\}$ only shows positive values, just as $c_{2}\{4\}$, which does not produce a real value of $v_{3}\{4\}$.}.  The fact that {\tt UrQMD} could not generate the experimentally observed triangular flow, together with the results shown Figs.~\ref{figure:c22}--\ref{fig:c3234}, strongly indicates that p--Pb systems from {\tt UrQMD} contain large non-flow azimuthal correlations.

\begin{figure*}[tbh]
\begin{center}
\includegraphics[width=0.9\textwidth]{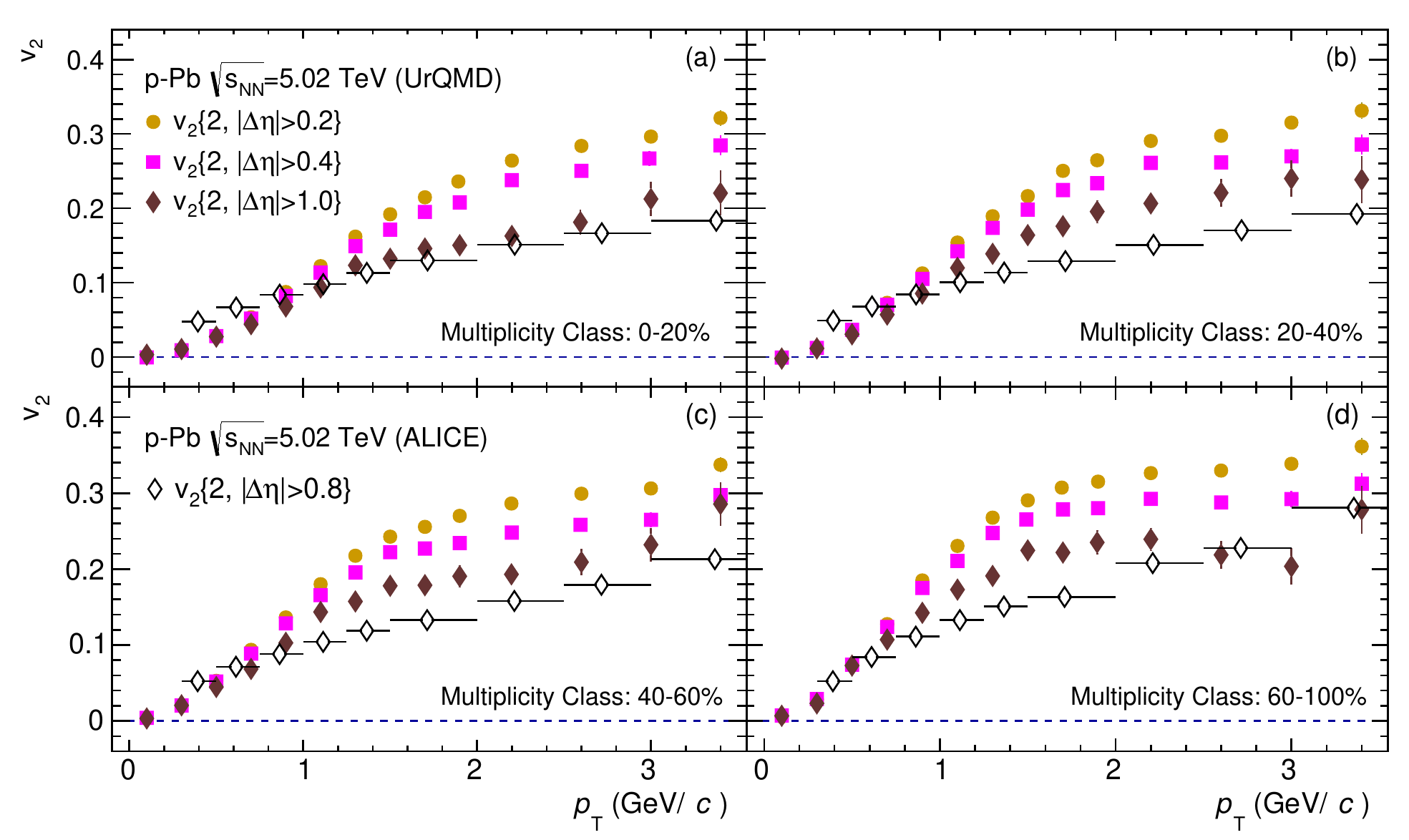}
\caption{(Color online) $v_{2}(\it{p}_{\rm T})$ of all change hadrons in p--Pb collisions at $\sqrt{s_{_{\rm NN}}} =$ 5.02 TeV, calculated from {\tt UrQMD} and measured by ALICE~\cite{ABELEV:2013wsa}. Here the multiplicity class determination in {\tt UrQMD} is based on Table~\ref{table2}.}
\label{fig:v2chGaps}
 \end{center}
\end{figure*}

\begin{figure*}[tbh]
\begin{center}
\includegraphics[width=0.95\textwidth]{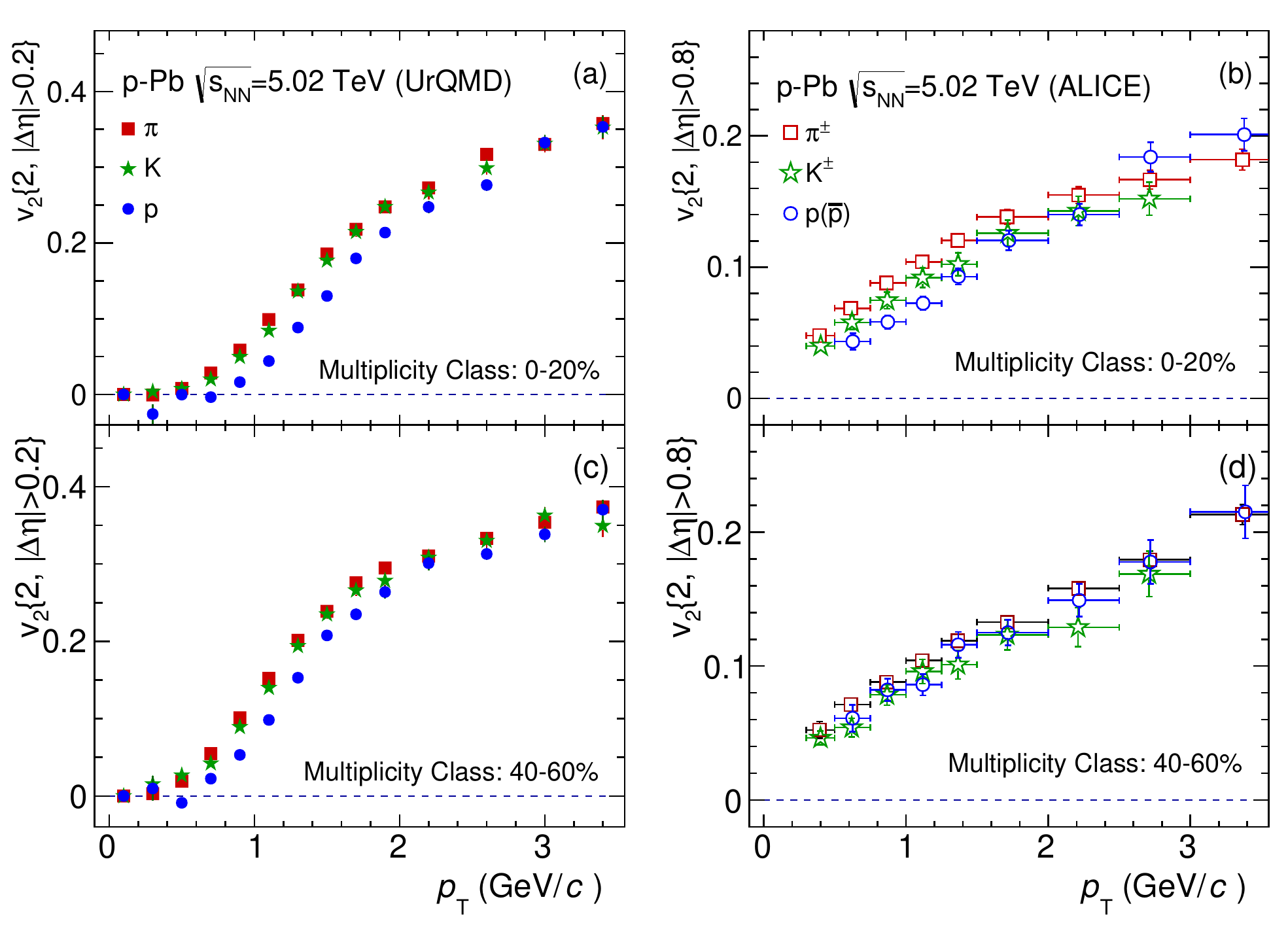}
\caption{(Color online) $v_{2}(\it{p}_{\rm T})$ of pions, kaons and protons in p--Pb collisions at $\sqrt{s_{_{\rm NN}}} =$ 5.02 TeV,  calculated from {\tt UrQMD}(left panels) and measured by ALICE(right panels)~\cite{ABELEV:2013wsa}. Here the multiplicity class determination in {\tt UrQMD} is based on Table~\ref{table2}.}
\label{fig:pidv2Gap02}
 \end{center}
\end{figure*}

Following Eq.~(\ref{vn2EtaGap}), we calculate the second Fourier flow-coefficient as a function of transverse momentum, $v_{2}(p_{\rm T})$, for the {\tt UrQMD} simulations at multiplicity class 0-20$\%$, 20-40$\%$, 40-60$\%$ and 60-100$\%$. Figure~\ref{fig:v2chGaps} shows that $v_{2}(p_{\rm T})$ monotonically increases from high to low multiplicity class, which agrees with the trend of $c_{2}\{2\}$ shown in Fig.~\ref{figure:c22} ($c_{2}\{2\}$ is the square of integrated $v_{2}\{2\}$). Meanwhile, $v_{2}(p_{\rm T})$ from {\tt UrQMD} increases with the increase of $p_{\rm T}$ and show strong sensitivity to the pseudorapidity gap.
The observed large pseudorapidity gap suppression of $v_{2}(p_{\rm T})$  indicates that non-flow effects are large in {\tt UrQMD}, as already shown in Fig.~\ref{figure:c22}-\ref{fig:c3234}.

Figure~\ref{fig:v2chGaps} also shows that {\tt UrQMD} can not correctly reproduce the shape of the experimental $v_2(p_{\rm T})$ curves, when implemented the same pseudorapidity gap $|\Delta \eta| > 1.0$. It underpredicts the data at lower $p_{\rm T}$ and overestimates the data above 1 GeV. Compared with the integrated $v_2$, the differential elliptic flow $v_2(p_{\rm T})$ contains more information on the evolving system, which reflects the interplay between radial and elliptic flow. The $m_{\rm T}$ spectra in Fig.~\ref{ptSpectra} has already shown that {\tt UrQMD} can not produce sufficient radial flow as observed in experiment. The insufficient radial flow, together with the insufficient flow anisotropy accumulation (shown in Fig.~3-5) leads to the fact that {\tt UrQMD} could not reproduced the $v_2(p_{\rm T})$ curves measured by ALICE.

Figure~\ref{fig:pidv2Gap02} investigates azimuthal correlations of identified hadrons in high energy p--Pb collisions. The right panels present the ALICE measurements with two different multiplicity classes ~\cite{ABELEV:2013wsa, Khachatryan:2014jra}, which show a characteristic feature of $v_{2}(p_{\rm T})$  mass-ordering among pions, kaons and protons. In the past research, hydrodynamic simulations from several groups have systematically studied the flow data, which reproduced the $v_2$  mass-ordering feature of the p--Pb systems~\cite{Bozek:2013ska, Werner:2013ipa}. In the hydrodynamic language, the radial flow further accumulated in hadronic stage tends to push heavier hadrons from lower $p_{\rm T}$ to higher $p_{\rm T}$, leading to an enhanced $v_2$ splitting between pions and protons~\cite{Hirano:2007ei,Song:2013qma}. The observation of $v_2$ mass-ordering is thus generally believed as a strong evidence for the collective expansion of the p-Pb systems created in $\sqrt{s_{_{\rm NN}}}=$ 5.02 TeV collisions.

However, the left panels of Fig.~\ref{fig:pidv2Gap02} shows that {\tt UrQMD} also generate a mass-ordering for the 2-particle correlations among pions, kaons and protons~\footnote{Due to limited statistics, we apply $|\Delta\eta|>0.2$ in our calculations, rather than $|\Delta\eta|>0.8$ as used in experiment.  In fact, $v_{2}\{2; |\Delta\eta|>0.8\}$ from our current {\tt UrQMD} simulations has large error bars, especially for protons. However, a tendency of $v_2$ mass ordering among pions, kaons and protons is still observed.}. Such mass-ordering pattern, caused by pure hadronic interactions, qualitatively agrees with the ones from the ALICE measurement~\cite{ABELEV:2013wsa} and from the hydrodynamic calculations~\cite{Bozek:2013ska, Werner:2013ipa}. In {\tt UrQMD}, the unknown cross sections are calculated by the additive quark model (AQM) through counting the number of constituent quarks within two colliding hadrons. As a result, the main meson-baryon (M-B) cross sections from AQM are about 50\% larger than the meson-meson (M-M) cross sections, leading to the  $v_{2}$ splitting between mesons and baryons after the evolution of hadronic matter.  Comparison simulations in appendix A (Fig.~9) also show that, with the M-B and M-M interaction channels closed in {\tt UrQMD}, the $v_{2}$ mass-ordering almost disappears. The combined results in Fig.~\ref{fig:pidv2Gap02} and~\ref{v2woMM} illustrate that the hadronic interactions could lead to a mass-ordering in 2-particle correlations among pions, kaons and protons, even for small p--Pb systems without enough flow generation.

\section{Summary }
\label{s:summary}

Using {\tt UrQMD} hadron cascade model, we studied azimuthal correlations in p--Pb collisions at $\sqrt{s_{_{\rm NN}}} =$ 5.02 TeV. Comparisons with the experimental data showed that the p-Pb systems created in experiment are not the trivial hadronic systems described by {\tt UrQMD}. Here, we summarize the main results as the following:

\textbf{(1)} With large pseudorapidity gaps ($|\Delta \eta|>1.0$), the measured 2-particle cumulant of the second Fourier flow-coefficient $c_{2}\{2\}$ from ALICE shows a weak centrality dependence from central to semi-peripheral collisions. In contrast, the {\tt UrQMD} calculations still present a strong centrality dependence for $c_{2}\{2, |\Delta \eta|>1.0\}$.

\textbf{(2)} In the most central collisions, $c_{2}\{4\}$ from ALICE exhibits a transition from positive to negative values, which indicates the development of strong collective flow in high multiplicity events. However, $c_{2}\{4\}$ from {\tt UrQMD} keeps positive for all multiplicity classes, which does not produce $v_2\{4\}$ with a real value.

\textbf{(3)} For large pseudorapidity gaps, $c_{3}\{2\}$ from {\tt UrQMD} turns to negative values, which can not produce the triangular flow as observed in experiments.

\textbf{(4)} {\tt UrQMD} can not fit the differential flow $v_2(p_{\rm T})$ from ALICE at various multiplicity classes.

\begin{figure*}[thb]
\includegraphics[width=0.95\textwidth]{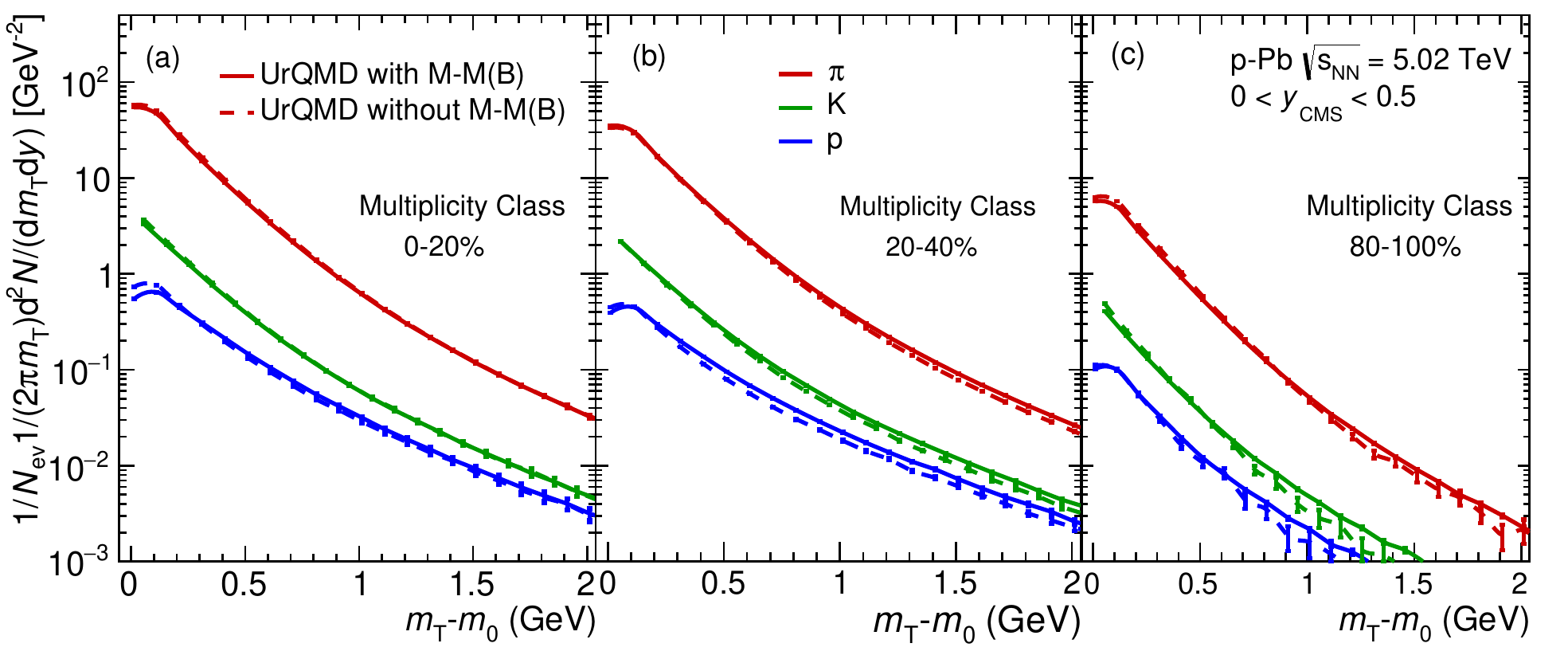}
\caption{$m_{\rm T}$ spectra of of pions, kaons and protons  in p--Pb collisions at $\sqrt{s_{_{\rm NN}}} =$ 5.02 TeV, calculated from {\tt UrQMD} with and without M-M and M-B collisions. Here the multiplicity class determination in {\tt UrQMD} is based on Table~\ref{table2}.}
\label{ptSpectra-2}
\end{figure*}

More specifically, the related experimental data of azimuthal correlations have accumulated strong evidence for the development of strong collective flow in high multiplicity events. With the assumption that high energy p-Pb collisions do not reach the threshold for the QGP formation and only produce trivial hadronic systems, we did hadron transport simulations with {\tt UrQMD}. We found that hadronic interactions alone could not generate sufficient collective flow as observed in experiment. Non-flow effects, e.g. from resonance decays and/or jet-like fragmentations, largely influence the hadron emissions of the {\tt UrQMD} systems. In order to fit the measured azimuthal correlations of all charged hadrons in p-Pb collisions at $\sqrt{s_{_{\rm NN}}} =$ 5.02 TeV, the contributions from the initial stage and/or the QGP phase can not be neglected.

In addition, we extended our study of azimuthal correlations to identified hadrons. The calculations of the 2-particle correlations for pions, kaons and protons showed that {\tt UrQMD} can generate a $v_2$ mass-ordering with the characteristic feature similar to the ALICE measurements.
Comparison runs from UrQMD with main hadronic scatterings turned on and off showed that the $v_2$ mass-ordering in UrQMD is mainly caused by hadronic interactions. The $v_2$ mass-ordering alone is not necessarily a flow signature associated with strong fluid-like expansions.

\begin{figure*}[thb]
\includegraphics[width=0.45\textwidth]{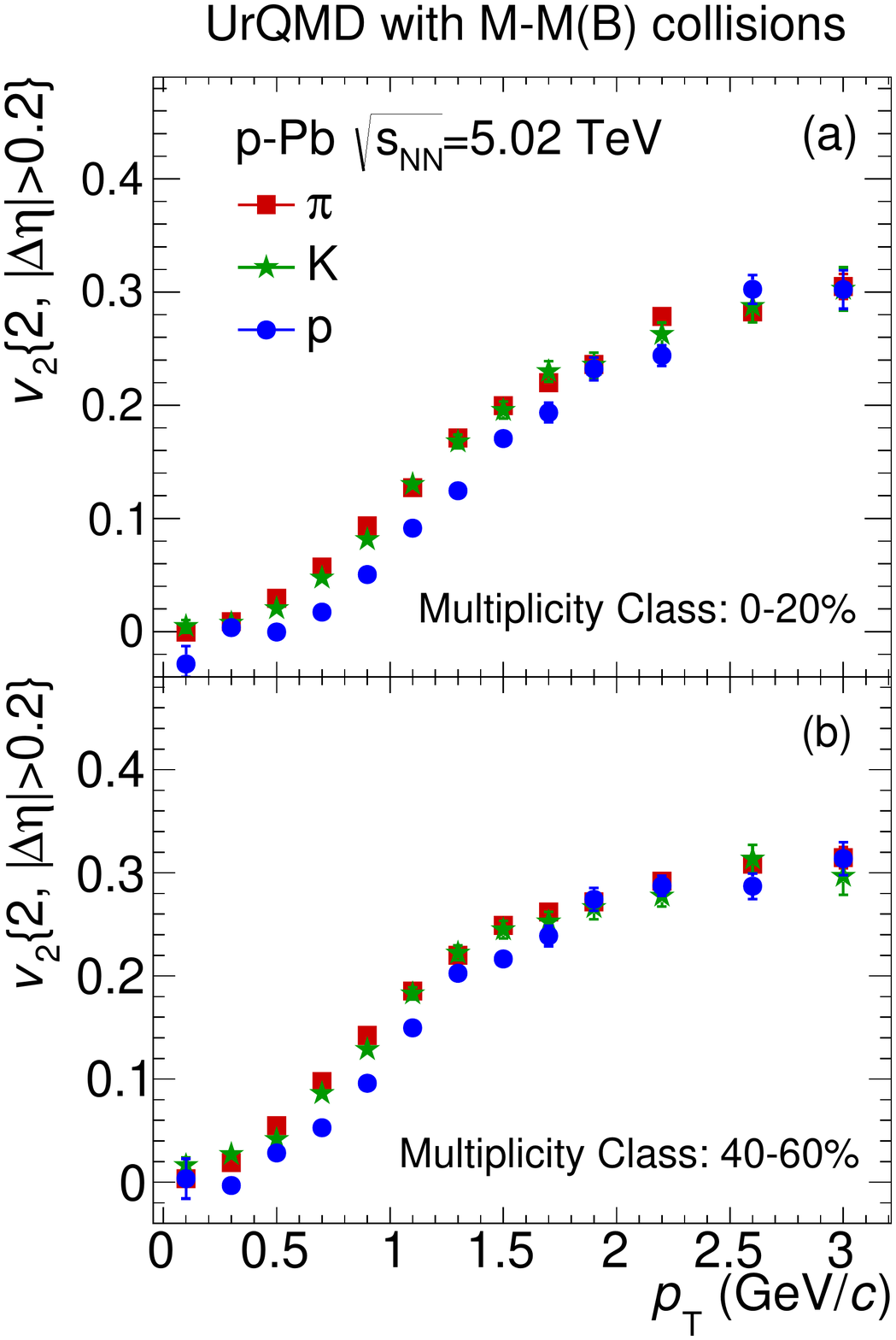}
\includegraphics[width=0.45\textwidth]{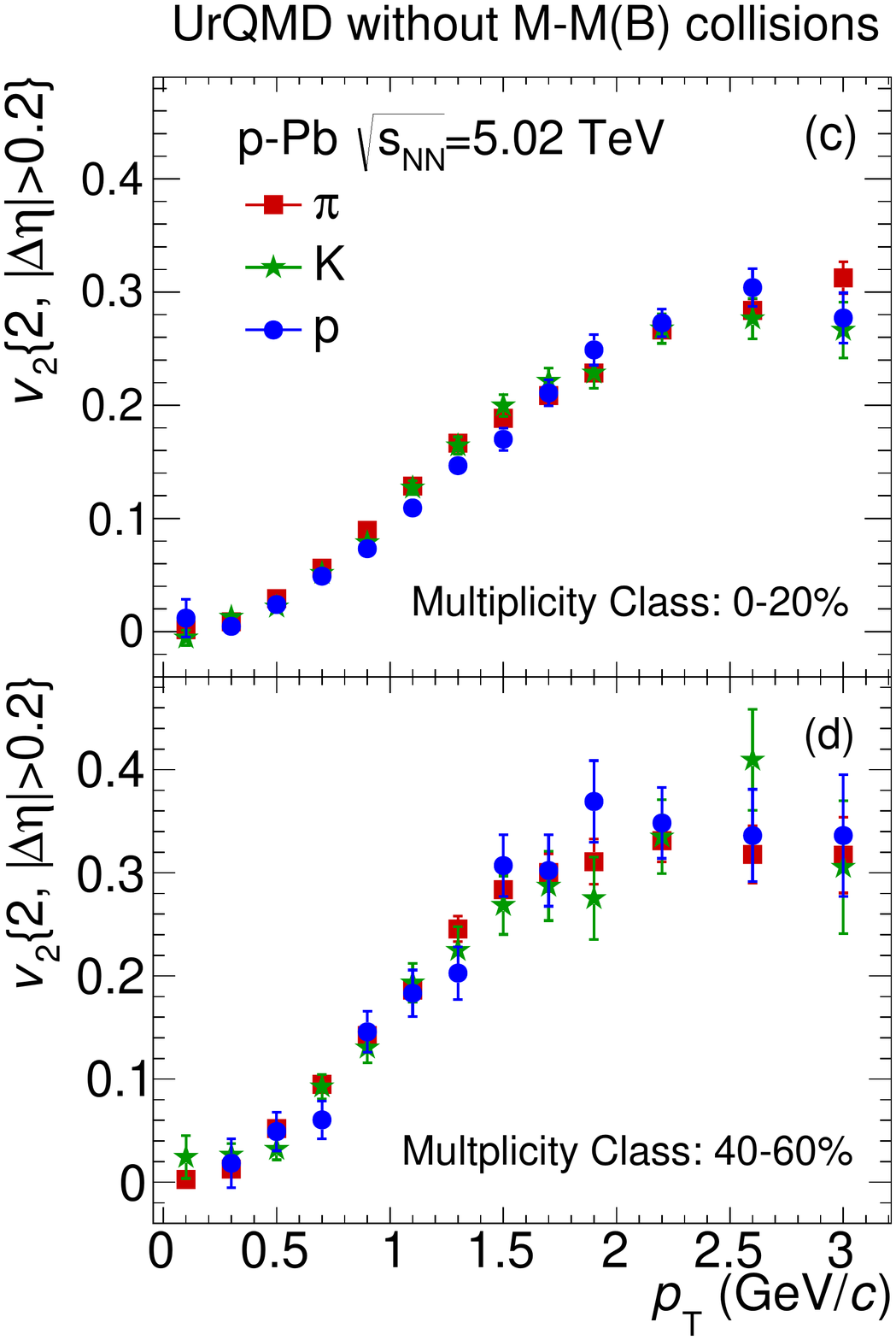}
\caption{ $v_{2}(\it{p}_{\rm T})$ of pions, kaons and protons in p--Pb collisions at $\sqrt{s_{_{\rm NN}}} =$ 5.02 TeV, calculated from {\tt UrQMD} with and without M-M and M-B collisions.
Here the multiplicity class determination in {\tt UrQMD} is based on Table~\ref{table2}. }
\label{v2woMM}
\end{figure*}

\acknowledgments
\vspace*{-2mm}
We thank S.~A.~Bass, A.\,Biland\v{z}i\'{c}, J.~J.~Gaardh\o je, U.~Heinz, J.~Y.~Ollitrault, H.~Petersen, J.~Schukraft and R.~Snellings for valuable discussions. 
YZ thanks to the Danish Council for Independent Research, Natural Sciences and the Danish National Research Foundation (Danmarks Grundforskningsfond) for support, thanks to Peking University for the host. XZ, PL and HS are supported by the NSFC and the MOST under grant Nos. 11435001 and 2015CB856900. We gratefully acknowledge extensive computing resources provided to us on Tianhe-1A by the National Supercomputing Center in Tianjin, China.

\appendix

\section{$\mathrm{\mathbf{UrQMD}}$ comparison runs with/without $\mathrm{\mathbf{M-M}}$ and $\mathrm{\mathbf{M-B}}$ collisions}

This appendix explores how hadronic interactions in {\tt UrQMD} influence spectra and azimuthal correlations of identified hadrons for the hadronic p--Pb systems. In {\tt UrQMD}, hadronic scatterings include Meson-Meson (M-M) collisions, Meson-Baryon (M-B) collisions and Baryon-Baryon (B-B) collisions. When switching off  all of these collision channels, {\tt UrQMD} simulations, in principle, consist of initial hadron productions and the succeeding resonance decays, which are mainly influenced by non-flow effects. However, not all these collision channels in the current version of {\tt UrQMD} (v3.4) can be simultaneously turned off. With B-B collision channels turned off, all the secondary proton-nucleon collisions from the initial p--Pb collisions are automatically turned off without proceeding any further hadron productions and decays.
Considering the probability of B-B collisions is much lower than M-M and M-B collisions, we only turn off the M-M and M-B interaction channels for the {\tt UrQMD} comparison runs in this appendix.

Figure~\ref{ptSpectra-2} plots the $m_{\rm T}$ spectra of pions, kaons and protons in p--Pb collisions at $\sqrt{s_{_{\rm NN}}} =$ 5.02 TeV, calculated from {\tt UrQMD} with and without M-M and M-B interactions. In Sec.~\ref{s:RandD}, we have already showed that, although the $m_{\rm T}$ scaling is weakly broken in {\tt UrQMD}, pure hadronic interactions can not generate sufficient radial flow as observed in experiment. Here, Fig.~\ref{ptSpectra-2} shows that M-M and M-B collisions only slightly change the slope of the $m_{\rm T}$ spectra~\footnote{For 2.76 A TeV Pb+Pb collisions at 70-80\% centrality (where $dN_{ch}/d\eta$ at mid-rapidity is about 50~\cite{Song:2013qma} and close to $dN_{ch}/d\eta$ in high multiplicity p--Pb collisions), hybrid model simulations also show that hadronic scatterings almost do not change the $m_{\rm T}$ spectra of identified hadrons~\cite{ZHU}.}. The slight broken of the $m_{\rm T}$ scaling  in {\tt UrQMD} are very possibly caused by mechanisms of the initial hadron productions~\footnote{In the {\tt UrQMD} simulations for p-Pb collisions at $\sqrt{s_{_{\rm NN}}} =$ 5.02 TeV, most of the initial hadron productions are triggered by the PYTHIA mode due to large momentum transfer~\cite{Bass:1998ca,Bleicher:1999xi,Petersen:2008kb,Bass}. Backup simulations from PYTHIA with collision energy set to 5.02 TeV show that the $m_{\rm T}$ scaling are also weakly broken for the $m_{\rm T}$ spectra of pions, kaons and protons~\cite{ZHU}.}.

Compared with the $m_{\rm T}$ spectra, the $v_2$ mass-ordering are more sensitive to hadronic interactions. For typical flow-dominated systems created in high-energy Au--Au or Pb--Pb collisions, hybrid model simulations have shown that hadronic rescatterings dramatically increase the $v_2$ splitting between pions and protons, but only slightly change the $m_{\rm T}$ spectra~\cite{Song:2010aq}.  Figure~\ref{v2woMM} presents $v_{2}(p_{\rm T})$ of identified hadrons in high energy p--Pb collisions, based on {\tt UrQMD} simulations in the senarios of with  (left panels) and without (right panels) M-M and M-B collisions. In the cases that M-M and M-B collisions are turned off, the $v_2$ mass-ordering among pions, kaons and protons almost disappears, when compared with the cases with M-M and M-B interactions.

In Sec.~IV, detailed study of the 2-particle and 4-particle correlations has already shown that {\tt UrQMD} could not generate sufficient flow as observed in experiments, its final hadron emissions are largely influenced by non-flow effects.  The comparisons runs in Fig.~9 illustrate that the $v_2$ mass-ordering can be explained as the consequence of hadronic interactions, which is not necessarily associated with strong fluid-like expansions.


\end{document}